\documentstyle[12pt]{article}

\catcode`@=11
\def\secteqno{\@addtoreset{equation}{section}%
\def\theequation{\thesection.\arabic{equation}}}
\catcode`@=12
\secteqno
\newcommand{\be}{\begin{equation}}
\newcommand{\ee}{\end{equation}}
\newcommand{\bea}{\begin{eqnarray}}
\newcommand{\eea}{\end{eqnarray}}
\newcommand{\bref}[1]{(\ref{#1})}

\catcode`@=11

\begin{document}
\vfill
\vbox{
\today
\hfill 
}\null

\vskip 10mm
\begin{center}
{\Large\bf Gravitational Field as a Generalized Gauge Field Revisited
\\
}\par
\vskip 10mm
{Takeshi FUKUYAMA\footnote{\tt E-mail:fukuyama@se.ritsumei.ac.jp}}\par
\medskip
Department of Physics and R-GIRO, Ritsumeikan University,\\ 
Kusatsu, Shiga, 525-8577 Japan\\
\medskip
\vskip 10mm
\end{center}
\vskip 10mm
\begin{abstract}
We add some comments to our old paper \cite{F-U} where the metric tensor was introduced as the gauge theory of general coordinate transformation.
This formulation is more satisfactorily completed than the original one if it is required to be valid for arbitrary n dimensional spacetime.
That is, our formulation asserts the presence of extra dimensions positively.
\end{abstract}


Almost 40 years ago we published the paper titled Gravitational Field as a Generalized Gauge Field \cite{F-U}, where the gravitational field was formulated as the gauge field of the general coordinate transformation.
In that paper, the dimensionality of spacetime was fixed as four.  Just after having published it, we noticed that this theory is more satisfactorily formulated than the original one if we required that the theory should be valid in any dimension.
This requirement imposes more stringent constraints on the theory and is sufficient for determing all free parameters introduced in the formulation. We have not published this version so far. 

However, nowadays, the expectation that new physics beyond the Standard Model may be realized by the presence of the extra dimensions is getting more and more realistic. 
The extra dimensions were discussed in the Kaluza-Klein \cite{Kalza} soon after the success of General Relativity and revived in its modern version of gauge-Higgs unfication \cite{Hosotani}.
The extra dimension also comes from SUGRA \cite{Wess} and superstring \cite{Green} and the consistency of GUT \cite{Kawamura}.

In these situations, it may be meanigful to publish the version, stating that gravitation needs extra dimensions from gauge theoretical construction without incorpolating the symmetry other than the general coordinates invariance.
In our knowledge, this standpoint still remained new except for ourselves.

The Reimannian tensor has the universal form irrelevant to the dimensionality,
\be
R^\mu{}_{\nu\rho\sigma}=\partial_\rho\Gamma^\mu_{\nu\sigma}-\partial_\sigma\Gamma^\mu_{\nu\rho}+\Gamma^\mu_{\alpha\rho}\Gamma^\alpha_{\nu\sigma}-\Gamma^\mu_{\alpha\sigma}\Gamma^\alpha_{\nu\rho},
\ee
where the world suffix runs from $1$ to $n$ in n dimensional spacetime.
The gravitational action has the same form of the Rieman scalar or the scalar composed of the several Riemannian tensors irrelevant to the dimensionalty.
However, they neither require the extra dimensionality from the model building
nor assert the presence of extra dimension.
On the other hand, SUGRA has the maximum dimesion 11 and Superstring has the consistent dimension 10. This is because the latters have more constraints, like supersymmetry and Lorentz invariance etc. other than the general covariance.

In this letter we emphasize that the gravitational field as the generalized gauge field is consistently formulated iff it is requsted to be consistent in any dimension.
This fact is derived by straightforwardly following \cite{F-U}.
In \cite{F-U}, we considered in four dimensions and the completion of formulation required a kind of solvability in addition to the gauge invariance constraints.
This defects is solved by requiring that theory should be valid in any dimension.
Formulation is not modified except for that the dimension is generalized from four to general n (n:integer). So the change comes only from the contraction like
\be
\delta_\mu^\mu=4 \rightarrow n.
\ee

So we do not repeat the argument of \cite{F-U} but write only the difference and result. Indeed, the change of formulation is very trivial, but the results are not.\\
Before starting the arguments, we comment on the generalised gauge field adopted in \cite{F-U} and this letter.
In the usual gauge principle, local gauge invariance is the invariance under
\be
\delta \phi^A\equiv iT_{(a)}^A{}_B\epsilon^a (x)\phi^B.
\label{tr}
\ee
for matter field $\phi^A$. Here $a$ is the group index. $T_{(a)}^A{}_B$ is the generator of gauge group and $\epsilon^a$ is the x-dependent magnitude of transformation in $a$-direction \cite{Utiyama2}.
So the number of group index $a$ is independent on the spacetime dimensionality. 
The gauge field and its transformation property id determined to
\be
\delta A_\mu^a=f_b{}^a{}_cA_\mu^b\epsilon^c+\partial_\mu\epsilon^a
\label{tf3}
\ee

For the case of gravity, the situation is a little bit changed. We can make vanish gravitation locally by equivalence principle. That is, the internal group is soldered to external world coordinates. Also the action is not quadratic in field strength (Riemannian tensor) but linear.

As one of the great contributions in his seminal work \cite{Utiyama2}, Utiyama incorporated spin connections as the gauge field of local Lorentz group but tetrad was treated as external fields. The Poincare group \cite{Kibble} includes tetrad but it is not clear why the tetrad transforms covariantly unlike \bref{tf3}. 
The (anti) de Sitter \cite{MacDowell} gauge group is most satisfactory in these sences. This gauge group invokes the other interesting connections with the usual gauge fields \cite{Maldacena} and discussed in separate form \cite{Fukuyama2}.

However, these are not all of the approaches. In this paper, the primary object is the metric tensor, and the Christoffel symbols are derived as the fuctions of the gauge field.

We consider the gauge field $A_K$ which is connected with the local homogeneous linear coordinate transformation,
\be
dx'^\mu=dx^\mu+\epsilon^\mu{}_\nu(x)dx^\nu,
\ee
which is integrated to
\be
x'^\mu=x^\mu+\xi^\mu (x)
\ee
under the integrability condition
\be
\frac{\partial}{\partial x^\sigma}\epsilon^\mu{}_\rho(x)=\frac{\partial}{\partial x^\rho}\epsilon^\mu{}_\sigma(x),
\ee
and therefore 
\be
\epsilon^\mu{}_\rho=\partial _\rho\xi^\mu\equiv \xi^\mu{}_{,\rho}.
\ee
In order to cancell not only $\partial \xi/\partial x$ but also $\partial^2 \xi/\partial x^2$, we we have two approaches as we have mentioned in \cite{F-U}.

(a) According to \bref{tf3}, we introduce the gauge field (the Christoffel symbols)
\be
\delta A^\rho{}_{\mu\nu}=\xi^\alpha{},_\beta\cdot \left({}^\beta{}_\alpha|N^\rho{}_{\mu\sigma}\right)-\xi^\rho{}_{,\mu\sigma},
\ee
where $\left({}^\beta{}_\alpha|N^\rho{}_{\mu\sigma}\right)$ should be determined from the gauge principle \cite{Utiyama}.
In this case the metric is treated as so called external field.

(b) Another is the approach sdopted in \cite{F-U} and here, requiring that Lagrangian depend upon $\partial A_K/\partial x$ in addition to $A_K$.
In this approach, the metric is the primary "gauge" field which is transformed as an irreducible tensor of r-th rank with full symmetry, $A_K=A_{(\mu_1...\mu_r)}$,
\bea
\delta \phi_A(x)&=&\phi_B\cdot C^{B\nu}_{A\mu}\cdot \xi^\mu{}_{,\nu}\\
\delta A_{(\mu_1...\mu_r)}(x)&=&A_{(\nu_1...\nu_r)}\cdot D\left(^{\nu_1...\nu_r}_{\mu_1...\mu_r}\right)^\nu_\mu\cdot \xi^\mu{}_{,\nu}.
\eea
Here $\phi_A$ is a kind of tensor. $C_{A\mu}^{B\nu}$ and $D\left(^{\nu_1...\nu_r}_{\mu_1...\mu_r}\right)^\nu_\mu$ is described as the products of Kronecker's deltas.
Nother's theorem shows that $\phi_{A,\lambda}$ and $A_{(\mu_1...\mu_r),\lambda}$ should be included in a particular linear combination \cite{F-U}
\be
\nabla_\lambda \phi_A=\partial_\lambda\phi+\phi_BM_{A\lambda}^{B\mu_1...\mu_r\nu}(x)A_{(\mu_1...\mu_r),\nu}.
\label{covariant}
\ee
Here $M_{A\lambda}^{B\mu_1...\mu_r\nu}(x)$ is described as a linear combination of $C_{A\alpha}^{B\beta}$,
\be
M_{A\lambda}^{B\mu_1...\mu_r\nu}(x)=C_{A\alpha}^{B\beta}Y_{\lambda\beta}^{\mu_1...\mu_r\nu\alpha}(x).
\ee
It should be remarked in \bref{covariant} that the derivative of $A_{(\mu_1...\mu_r)}$ appears in the covariant derivative. This gives a partial reasoning why action must be linear in the fieldstrength (Riemannian tensor). Usual square term of Riemannian tensor includes higher derivative terms of primary gauge field $A_{(\mu_1...\mu_r)}$, which breaks unitarity in quantum theory.

If we define $A^{(\mu_1...\mu_r)}$ by
\be
A^{(\mu_1...\mu_r)}\propto Y_{\alpha\beta}^{\mu_1...\mu_{r-1}\alpha\beta\mu_r},
\ee
it satisfies
\be
A^{(\mu_1...\mu_{r-1}\alpha)}A_{(\mu_1...\mu_{r-1}\beta)}\propto \delta_\beta^\alpha.
\ee
We have reviewed the fundamental properties of \cite{F-U} (Please see the original paper for the detail.) and proceed to the main theme of this letter.
\vskip 10 mm
Our arguments start from "the Determination of $A_{(\alpha_1...\alpha r)}$" of Appendix B in \cite{F-U}. All notations and conventions follow \cite{F-U}.
\bea
2\delta_\rho^\alpha\delta_{(\lambda\beta)}^{(\mu\nu)}&\equiv& r\cdot a\left[A_\rho^\mu\cdot \delta_{(\lambda\beta)}^{(\nu\alpha)}+A_\rho^\nu\cdot \delta_{(\lambda\beta)}^{(\mu\alpha)}\right]\nonumber\\
 &+&r\cdot b\left[2A_\rho^\nu\cdot \delta_{(\lambda\beta)}^{(\alpha\mu)}+2A_\rho^\mu\cdot \delta_{(\lambda\beta)}^{(\alpha\nu)}+2(r-1)\{\delta_\beta^\alpha A_{\lambda\rho}^{\mu\nu}+\delta_\lambda^\alpha A_{\beta\rho}^{\mu\nu}\}\right]\nonumber\\
&+&r\cdot c\left[4A_\rho^\alpha\delta_{(\lambda\beta)}^{(\mu\nu)}+(r-1)\{\delta_\beta^\nu A_{\lambda\rho}^{\mu\alpha}+\delta_\lambda^\nu A_{\beta\rho}^{\mu\alpha}+\delta_\beta^\mu A_{\lambda\rho}^{\nu\alpha}+\delta_\lambda^\mu A_{\beta\rho}^{\nu\alpha}\}\right]\nonumber\\
&+&r(r-1)\cdot d\left[2(r-2)\cdot A_{\lambda\beta\rho}^{\mu\nu\alpha}+\delta_\beta^\mu A_{\lambda\rho}^{\nu\alpha}+\delta_\lambda^\mu A_{\beta\rho}^{\nu\alpha}+\delta_\beta^\nu A_{\lambda\rho}^{\mu\alpha}+\delta_\lambda^\nu A_{\beta\rho}^{\mu\alpha}\right].
\label{contraction}
\eea
Here $r(>1)$ is the rank of $A_{(\alpha_1...\alpha_r)}$ and $a,...,d$ are parameters introduced by (3.9) of \cite{F-U}, all of which are to be determined from the gauge principle. Also the following abbreviations have been employed:
\be
A_\rho^\alpha\equiv A^{(\alpha\mu_2...\mu_r)}A_{(\rho\mu_2...\mu_r)},~~A^{\alpha\nu}_{\beta\rho}\equiv A^{(\alpha\nu\mu_3...\mu_r)}A_{(\beta\rho\mu_3...\mu_r)}~~\mbox{etc.}
\ee

The double contraction of \bref{contraction} by putting $\mu=\lambda$ and $\nu=\beta$ leads to
\be
\delta_\rho^\alpha n(n+1)=A_\rho^\alpha\left[ra(1+n)+2rb(n+2r-1)+2rc(n+1)(n+r-1)+2r(r-1)d(r+n-1)\right].
\ee
First $A_\rho^\alpha\propto \delta _\rho^\alpha$ and we can normalize 
\be
A_\rho^\alpha=\delta_\rho^\alpha
\label{A}
\ee
without losing generality, and we obtain the polynomial equation with respect to n.
If we require that this equation should be valid in any n dimensions, the coefficients of every power of n must be zero and then 
\bea
0&=&ra+2rb(2r-1)+2rc(r-1)+2r(r-1)^2d,\\
1&=&ra+2rb+2rc(r-1)+2r(r-1)d,\\
1&=&2rc.
\eea
In a similar way, the contraction of \bref{contraction} by putting $\alpha=\rho$ and $\mu=\lambda$ leads to
\bea
&&A\delta_\beta^\nu\left[\frac{(n+1)}{2}ra+rb(n+1)+rc(r+1)+r(r-1)d\right]\nonumber\\
&+&A_\beta^\rho\left[\frac{(n+1)}{2}ra+rb(n+1)(2r-1)+rc(3r-1)+r(r-1)d(2r-1)\right]\nonumber\\
&=&(n+1)\delta_\beta^\rho.
\eea
Here $A\equiv A_\rho^\rho=n$ from \bref{A}.
So we obtain
\bea
0&=&\frac{ra}{2}+rb,\\
1&=&\frac{ra}{2}+2r^2b+rc(r+1)+r(r-1)d,\\
1&=&\frac{ra}{2}+rb(2r-1)+rc(3r-1)+r(r-1)(2r-1)d.
\eea
So the above 6 constaraints are sufficient for the determination of the five parameters and we obtain uniquely
\be
a=b=0,~c=-d=\frac{1}{4}, r=2.
\label{parameter}
\ee
Thus we can derive the second ranked gauge field and the Christoffel symbols ((3.13) in \cite{F-U}) unambiguously,
\be
\nabla_\lambda\phi_A=\partial_\lambda\phi_A+\phi_B\cdot C^{B\beta}_{A\alpha}\cdot \Gamma^\alpha_{\beta\lambda}.
\ee
Here
\be
C^{A\alpha}_{B\beta}\equiv (A|{\bf C}^\alpha_\beta|B)
\ee
satisfies the commutation relation,
\be
[{\bf C}^\alpha_\beta,{\bf C}^\mu_\rho ]=\delta^\mu_\beta{\bf C}^\alpha_\rho-\delta^\alpha_\rho{\bf C}^\mu_\beta,
\ee
and
\be
\Gamma^\alpha_{\beta\lambda}=\frac{1}{2}A^{\alpha\sigma}\{A_{\sigma\lambda,\beta}+A_{\beta\sigma,\lambda}-A_{\lambda\beta,\sigma}\}.
\ee

The other types of contraction give the the same constraints as the above series of equations and conssitency is assured.
This is contrasted with the situation where the dimension is fixed as four.
In that case we could not determine the parameter like \bref{parameter}
unless we required the consistency condition additionally.
That is, unless we imposed the vanishing of some coeeficients undetermined from gauge principle, we were led to unsolvable relation
\be
A^{\alpha\nu}(x)\cdot A_{\beta\rho}(x)=\frac{2}{5}\delta_{\beta\rho}^{\alpha\nu}.
\ee

However, the very merit of invariance principle comes from the fact that it leads us to the determination of the Lagrangian form uniquely.
So the requirement that General Relativity must be formulated in any dimension is very important. That is, from this gauge theoretical viepoint
\be
S_G^{4+D}=\frac{1}{2}M_D^{2+D}\int d^4x\int d^Dy\sqrt{-g_{4+D}}R_{4+D}
\ee
is not only assured but also needed to exist for any extra D dimensions unless the other symmetry or constraint is imposed.

\section*{Acknowledgements}
The formulation developed in this letter is heavily due to the deceased Professor R.Utiyama though the present author is responsible for this addendum.
This work is supported in part by the grant-in-Aid 
for Scientific Research from the Ministry of Education, 
Science and Culture of Japan (No. 20540282).

\end{document}